\documentclass[pra,aps,reprint,superscriptaddress]{revtex4-2}

\usepackage{amsfonts,amssymb,amsmath}

\usepackage{newtxtext} 
\usepackage{newtxmath} 

\usepackage[capitalize]{cleveref}
\usepackage{multirow}
\usepackage{graphicx}
\usepackage{xcolor}
\usepackage{bm}
\usepackage{xspace,soul}

\begin{document}
\title{Massive particle interferometry with lattice solitons: robustness against ionization}

\author{Piero Naldesi}
\affiliation{Universit\'{e} Grenoble-Alpes, LPMMC, F-38000 Grenoble, France and CNRS, LPMMC, F-38000 Grenoble, France}
\affiliation{Institute for Quantum Optics and Quantum Information of the Austrian Academy of Sciences, Innsbruck, Austria}

\author{Peter D Drummond}
\affiliation{Centre for Quantum Science and Technology Theory,
Swinburne University of Technology, Melbourne 3122, Australia}

\author{Vanja Dunjko}
\affiliation{Department of Physics, University of Massachusetts Boston, Boston Massachusetts 02125, USA}

\author{Anna Minguzzi}
\affiliation{Universit\'{e} Grenoble-Alpes, LPMMC, F-38000 Grenoble, France and CNRS, LPMMC, F-38000 Grenoble, France}

\author{Maxim Olshanii}
\email{maxim.olchanyi@umb.edu}
\affiliation{Department of Physics, University of Massachusetts Boston, Boston Massachusetts 02125, USA}

\date{\today}

\begin{abstract}
We revisit the proposal of Castin and Weiss [Phys. Rev. Lett. vol. 102, 010403 (2009)] for using the scattering of a quantum matter-wave soliton on a barrier in order to create a coherent superposition state of the soliton being entirely to the left of the barrier and being entirely to the right of the barrier. In that proposal, is was assumed that the scattering is perfectly elastic, i.e.\ that the center-of-mass kinetic energy of the soliton is lower than the chemical potential of the soliton. Here we relax this assumption. Also, we introduce an interferometric scheme, which uses interference of soltions, that can be used to detect the degree of coherence between the reflected and transmitted part of the soliton. Using exact diagonalization, we numerically simulate a complete interferometric cycle for a soliton consisting of six atoms. We find that the interferometric fringes persist even when the center-of-mass kinetic energy of the soliton is above the energy needed for complete dissociation of the soliton into constituent atoms.  
\end{abstract}

\maketitle

\tableofcontents

\section{Introduction \label{sec_introduction}}
There is considerable interest in developing matter-wave interferometry with bright solitons, because they have a potental to improve sensitivity by as much as a factor of 100 \cite{cuevas2013_063006}, as well as enable high-precision force sensing \cite{cuevas2013_063006} and measurement of small magnetic-field gradients \cite{Veretenov2007_455}. A related possible application is the creation of Schr{\"o}dinger-cat states \cite{streltsov2009_043616,Streltsov2009_091004,weiss2009_010403}. Atomtronic devices with featuring soliton interferometry were considered in Refs.~\cite{naldesi2020enhancing,Polo2021_015015,naldesi2018_053001}. How to realize very narrow barriers to be used as beam splitters in soliton interfermoetry was explored in Ref.~\cite{grimshaw2021soliton}. What effect harmonic confinement can have on the internal degrees of freedom of a quantum soliton was investigated in \cite{Holdaway2012_053618}.  

The major inspiration for the present paper is the work of Castin and Weiss \cite{weiss2009_010403,weiss2012_455306}.  In that proposal, a bright soliton is scattered off a broad Gaussian barrier in such a way that after the scattering, the atomic gas is in a coherent superposition of being entirely to the left of the barrier and being entirely to the right of the barrier. Here is how that happens.

First of all, the parameter regime is such that if the soliton were a point particle, there would be a threshold for incoming velocity such that, if the incoming velocity is below the threshold, the particle is completely reflected, and if above, it is perfectly transmitted. The center-of-mass (CoM) velocity of the incoming soliton is tuned to be exactly at this threshold.

At the same time, the CoM soliton speed is kept low enough that the CoM kinetic energy of the soliton is lower than the chemical potential of the soliton. This ensures that the scattering is perfectly elastic:  the soliton is energetically forbidden from `ionizing' (shedding particles). 

Now we note that the soliton wavepacket is not perfectly monochromatic, and thus it contains Fourier components both slightly below and slightly above the threshold needed to cross the barrier. So the scattering process behaves as filtering in Fourier space, where the components that are below the threshold are completely reflected and those above it are completely transmitted. But, as we saw above, the soliton cannot fragment. Therefore, the only option for the soliton is to coherently split into a reflected and a transmitted part.

We should note that scattering of bright solitons on barriers such that the soliton is typically either wholly transmited or wholly reflected was experimentally realized and studied in \cite{Boisse2017_10007}. The authors say that their solitons are too large to form mesoscopic quantum superpositions in the process, but note that such superpositions should be obervable for smaller numbers of atoms.

The purpose of the present work is to investigate what happens if the assumption of perfectly ellastic scattering is relaxed, and the CoM kinetic energy is allowed to be above the ionization threshold. Moreover, our barrier, rather than being very broad, will be very narrow. 

To do this, we need to be able to detect the degree of coherence between the reflected and transmitted part of the soliton (or, rather, of what remains of the soliton after it possibly sheds some particles). For this purpose, we use the coherent splitting described above to construct an interferometer that is sensitive to the presence of an external constant field. The quality of the interferometric fringes is our measure of coherence.

Our main result is that, remarkably, the interferometric fringes persist even if the CoM kinetic energy of our six-particle soliton is high enough that it would be energetically allowed for the soliton to completely disintegrate upon impact with the barrier.

\section{Description of the model system \label{sec_model_system}}

\subsection{The lattice Hamiltonian used in the numerical calculations \label{ssec_lattice_H}}
We are considering the Hamiltonian
%
\begin{multline}
\hat{\mathcal{H}} = -J \sum_{j=1}^{L} (\hat{a}^{\dagger}_{j} \hat{a}^{}_{j+1} +\text{h.c.} ) + \frac{U}{2} \sum_{j=1}^{L}  \hat{n}_{j}(\hat{n}_{j}-1) 
 \\+ W \hat{n}_{j_{\text{center}}} +  \sum_{j=1}^{L}  \kappa_{\text{mirror}} (j-j_{\text{center}})^2 d^2\,  \hat{n}_{j} 
 \\+  \sum_{j=1}^{L}   (j-j_{\text{center}})\, F \,d \, \hat{n}_{j}\,.
\label{H_Bose-Hubbard+barrier}
\end{multline}
%
Here $\hat{n}_{j} \equiv  \hat{a}^{\dagger}_{j} \hat{a}^{}_{j}$, $J$ is the hopping constant, $U$ is the onsite coupling constant,  $W$ is the barrier strength, $\kappa_{\text{mirror}}$ is the spring constant for a harmonic potential used to form the interferometer arms, and $F$ is a constant force acting to the left---the ``phase object''.. We impose periodic boundary conditions, $\hat{a}^{}_{L+1} \equiv  \hat{a}^{}_{1}$.  We will be working in a sector with a fixed number of particles
 $\hat{N} \equiv \sum_{j=1}^{L} \hat{n}_{j}$.
We assume that number of sites $L$ is odd. The center point of the lattice is given by 
$j_{\text{center}} \equiv  \frac{L+1}{2} \,\,.$
The length 
$d$ is the distance between neighboring lattice sites.  

The initial state is the ground state of the following Hamiltonian:
\begin{multline*}
\hat{\mathcal{H}}_{0-} = -J \sum_{j=1}^{L} (\hat{a}^{\dagger}_{j} \hat{a}^{}_{j+1} +\text{h.c.} ) + \frac{U}{2} \sum_{j=1}^{L}  \hat{n}_{j}(\hat{n}_{j}-1) 
 \\+  \sum_{j=1}^{L}  \kappa_{\text{preparation}} (j-(j_{\text{center}}-L_{0}))^2 d^2 \,  \hat{n}_{j} 
\,.
\end{multline*}
The Hamiltonian $\hat{\mathcal{H}}_{0-}$ is similar to $\hat{\mathcal{H}}$ in Eq.~(\ref{H_Bose-Hubbard+barrier}). However,  $\hat{\mathcal{H}}_{0-}$ includes no barrier and no force. Furthermore, its harmonic potential has a different curvature and is offset to the left from the center by $L_{0}$ lattice sites.

\subsection{The auxiliary continuum Hamiltonian \label{ssec_continuum_H}}
We further introduce the Hamiltonian
\begin{multline}
\hat{\mathcal{H}} = \int_{-\infty}^{\infty} \!dx \, 
\left\{ 
\frac{\hbar^2}{2m}
\hat{\Psi}^{\dagger}_{x}\hat{\Psi}^{}_{x} 
\right.
\\
+
\frac{g}{2} \hat{\Psi}^{\dagger}\hat{\Psi}^{\dagger}\hat{\Psi}^{}\hat{\Psi}^{}
+
 \hat{\Psi}^{\dagger}\hat{\Psi}^{} V_{\text{barrier}}(x) 
\\
\left.
+ 
 \hat{\Psi}^{\dagger}\hat{\Psi}^{}  \frac{m \omega_{\text{mirror}}^2 x^2}{2}
+
 \hat{\Psi}^{\dagger}\hat{\Psi}^{}  F x
\right\}
\label{H_continuum}
\,,
\end{multline}
where 
$V_{\text{barrier}}(x)  = \tilde{g}\, \delta(x)$ and $\omega_{\text{mirror}}$ is the frequency of the harmonic potential that is used to form an interferometer.  The number-of-particles operator is $\hat{N} \equiv \int \!dx \, \hat{\Psi}^{\dagger}_{x}\hat{\Psi}^{}_{x}$. 

The initial state is the ground state of 
\begin{multline}
\hat{\mathcal{H}}_{0-} = \int \!dx \, 
\left\{ 
\frac{\hbar^2}{2m} \hat{\Psi}^{\dagger}_{x}\hat{\Psi}^{}_{x} 
+
\frac{g}{2} \hat{\Psi}^{\dagger}\hat{\Psi}^{\dagger}\hat{\Psi}^{}\hat{\Psi}^{}
\right.
\\
\left.
+ 
 \hat{\Psi}^{\dagger}\hat{\Psi}^{}  \frac{m \omega_{\text{preparation}}^2 (x-(-L_{0}d))^2}{2}
\right\}
\,.
\end{multline}
%

%
\subsection{Discrete-to-continuum map \label{ssec_discrete_to_continuum_UllJ}}
Effective one-body mass is given by
\begin{align*}
&
m = \frac{1}{2} \frac{\hbar^2}{Jd^2}
\,\,. 
\end{align*}
Next, let us introduce a two-body scattering length $a$, which is the same in both the continuum and the lattice cases:
\begin{align*}
&
a = -\frac{2 d J^{\text{rel.}}_{K=0}}{U}\left(1- \frac{1}{\pi^2} \frac{U}{J^{\text{rel.}}_{K=0}}\right)
\,\,. 
\end{align*}
The corresponding coupling constant is given in terms of the scattering length in the usual way, from which we may deduce its dependence on the parameters of the lattice model:
\begin{align*}
&
g = -\frac{\hbar^2}{(m/2)a} \approx U d
\,\,. 
\end{align*}
Here, $J^{\text{rel.}}_{K} = 2J\cos(Kd/2)$ is the hopping constant of the lattice on which the relative motion \cite{valiente2008_161002}.
For an explanation of the 
lattice renormalization factor $\left(1- \frac{1}{\pi^2} \frac{U}{J^{\text{rel.}}_{K=0}}\right)$, see Ref.~\cite{castin2004_89}.

Note that it is not an accident that the lattice and continuum models share the 
same scattering length. In fact, the effective continuum coupling $g$ is introduced in such a way that it reproduces 
the lattice scattering length exactly. 

In a similar manner, we introduce a scattering length for the particle-barrier interaction, 
\begin{align*}
&
\tilde{a} = -\frac{2 d J}{W}\left(1- \frac{1}{\pi^2} \frac{W}{J}\right)
\,,
\end{align*}
and the corresponding particle-barrier coupling constant:
\begin{align*}
&
\tilde{g} = -\frac{\hbar^2}{m\tilde{a}} \approx Wd
\,\,.
\end{align*}

The frequency of the ``mirror'' satisfies 
\(
\frac{1}{2}m \omega_{\text{mirror}}^2 = \kappa_{\text{mirror}}
\),
so that
\[
\omega_{\text{mirror}} = \frac{2 d}{\hbar}\sqrt{\kappa_{\text{mirror}} J}
\,.
\]

Analogously, the ``preparation'' frequency will be given by 
$
\omega_{\text{preparation}} = \frac{2 d}{\hbar}\sqrt{\kappa_{\text{preparation}} J}
$.

\section{The regime of interest: no lattice renormalization and the continuum model applies \label{sec_weak_coupling}}
%
%
%
We assume that there is no need for lattice renormalization, either for the interactions or for the barrier. This happens in the regime of weak coupling, 
\[|U| \ll 2 \pi^2 J \quad\text{and}\quad W \ll  \pi^2 J\,.\]

Further, we assume that the continuum model applies, which will be the case when the healing length
\begin{align}
&
\ell =  2 \, \frac{\hbar^2}{m|g|N} = \frac{a}{N}
\label{healing_length}
\end{align}
satisfies $\ell \gg d$. This is equivalent to 
\[UN \ll 4 J\,.\]
%

\subsection{Preparation}  The initial soliton will be prepared so that its center of mass (CoM) is in the ground state of a ``preparation'' harmonic trap, of frequency $\omega_{\text{preparation}}$. Consider the mean kinetic energy of the soliton,
\begin{equation}
  E_{\text{kinetic, CoM}} =\frac{1}{2} M \bar{\mathcal{V}}^{2},
  \label{CoM_kinetic_energy}
\end{equation}
where $M = N m$ is the soliton mass and $\bar{\mathcal{V}}$ is the mean CoM velocity. Also consider the uncertainty in this kinetic energy
\begin{align*}
&
\delta E_{\text{kinetic, CoM}} \approx M \bar{\mathcal{V}} \delta\mathcal{V}
\,\,,
\end{align*}
where 
\begin{align*}
&
\delta\mathcal{V} = \sqrt{\frac{\hbar \omega_{\text{preparation}}}{2M}}
\end{align*}
is the preparation r.m.s.\ velocity. We will be assuming that the uncertainty in the kinetic energy is finite, but small relative to the mean kinetic energy. It then follows that the preparation r.m.s.\ velocity is small relative to the mean CoM velocity:
\begin{align*}
&
\delta E_{\text{kinetic, CoM}} \ll E_{\text{kinetic, CoM}} \Rightarrow \delta\mathcal{V} \ll \bar{\mathcal{V}}
\,.
\end{align*}

\subsection{Beam-splitting} The condition for a  50\%--50\%  classical filtering as quantum beam-splitting reads
\cite{weiss2009_010403,weiss2012_455306}
\begin{align*}
&
E_{\text{kinetic, CoM}}  = \max_{X}V_{\text{soliton-on-barrier}}(X)
\,,
\end{align*}
where $X$ is the CoM position. Note that when the number of atoms $N$ is large, the right-hand side is given by Eq.~(\ref{barrier_top}), below. See Ref.~\cite{castin2009_317} for a clear derivation of the one-body density matrix for a localized CoM position. The original result is in 
Ref.~\cite{calogero1975_265}.

\subsection{Mirrors and recombination}.
We will be using another harmonic trap, of frequency $\omega_{\text{mirror}}$, as a ``mirror'' on each end, to ensure the return of the wavepackets for recombination. The trap frequency and the 
initial position of the CoM wavepacket, $-L_{0}$, will conspire to produce the incident energy we need:
\begin{align}
&
\frac{M \omega_{\text{mirror}}^2 (L_{0})^2 }{2} =  \max_{X}V_{\text{soliton-on-barrier}}(X)
\,,
\label{trap_frequency_to_incident_E}
\end{align}
where, again, if the number of atoms $N$ is large, the right-hand side is given by Eq.~(\ref{barrier_top}).

\subsection{Inelastic effects in beam-splitting}.
\label{sec:inelastic_effects}
The full internal energy of the soliton, in the continuum limit (C.\ L.), is given by
\begin{align}
&
E^{(N)}_{\text{S, interaction}} \stackrel{\text{C.\ L.}}{=} -  \, \frac{(N+1)N(N-1)}{24}  \frac{mg^2}{\hbar^2} 
\label{soliton_energy_continuous_UllJ}
\,.
\end{align}
The gap to the first excitation is 
\begin{align*}
&
E^{(N-1)}_{\text{S, interaction}} - E^{(N)}_{\text{S, interaction}} \stackrel{\text{C.\ L.}}{=}  \frac{N(N-1)}{8} \frac{mg^2}{\hbar^2} 
\,;
\end{align*}
the gap to the first two-atom excitation is 
\begin{align*}
&
E^{(N-2)}_{\text{S, interaction}} - E^{(N)}_{\text{S, interaction}} \stackrel{\text{C.\ L.}}{=}  \frac{(N-1)^2}{4} \frac{mg^2}{\hbar^2} 
\,.;
\end{align*}
the gap to the first three-atom excitation is 
\begin{align*}
&
E^{(N-3)}_{\text{S, interaction}} - E^{(N)}_{\text{S, interaction}} \stackrel{\text{C.\ L.}}{=}  \frac{3N^2-9N+8}{8} \frac{mg^2}{\hbar^2} 
\,;
\end{align*}
and the gap to a complete `ionization' is 
\begin{align*}
&
0 - E^{(N)}_{\text{S, interaction}} \stackrel{\text{C.\ L.}}{=}   \frac{(N+1)N(N-1)}{24}  \frac{mg^2}{\hbar^2} 
\,. 
\end{align*}

We will be studying how the stability of the CoM interferometric signal depends on the potential for ionization \footnote{It may seem that breathing excitations constitute another potential inelastic channel. However, curiously, the soliton does not 
possess true localized excitations: the mean-field breathers are, in reality, unbound. Interestingly, the absence of bound excitations is 
confirmed in the Bogoliubov approximation. Such a restoration may seem accidental, if Bogoliubov is to be considered as an approximation 
of mean-filed equations. However, as a linearization of the Heisenberg equations of motion for the quantum field, Bogoliubov can give predictions that are more accurate than the mean-field ones, albeit limited to (perhaps multiple) monomer excitations.}. Note the following: (a) Refs.~\cite{weiss2009_010403} and \cite{weiss2012_455306}
conjecture that being below the single-atom ionization threshold is a prerequisite for a coherence between  the transmitted and reflected parts of the CoM wavepacket;
(b) Ref.~\cite{streltsov2009_043616} indicates that such coherence is preserved even above the six-atom ionization threshold, in a 100-atom soliton. 

Thus, the ionization threshold is
\begin{align}
&
E_{\text{kinetic, CoM}} > E^{(N-1)}_{\text{S, interaction}} - E^{(N)}_{\text{S, interaction}}\,.
\nonumber
\intertext{This works out to} 
&
E_{\text{kinetic, CoM}}  > \frac{N(N-1)}{8} \frac{mg^2}{\hbar^2} 
\label{exact_ionization_threshold}
\,.
\end{align}
%
%
Meanwhile, the following gives the energy window in which a one-atom ionization is allowed but already a two-atom one is forbidden:
\begin{multline*}
 E^{(N-2)}_{\text{S, interaction}} - E^{(N)}_{\text{S, interaction}} \ge
 \\
 E_{\text{kinetic, CoM}} > E^{(N-1)}_{\text{S, interaction}} - E^{(N)}_{\text{S, interaction}}\,,
\end{multline*}

%
%
which works out to
\begin{align}
& \frac{(N-1)^2}{4} \frac{mg^2}{\hbar^2}  \ge E_{\text{kinetic, CoM}}  > \frac{N(N-1)}{8} \frac{mg^2}{\hbar^2} 
\label{exact_single_ionization_window}
\,.
\end{align}
\subsection[Calculations for $N\gg 1$]{Calculations for $\bm{N\gg 1}$}
If the number of atoms is large, the mean-field approximation simplifies the calculations. 

The number-density distribution in a soliton is
\begin{align*}
&
n(x) = \frac{1}{2} \frac{N}{\ell}  \text{sech}^2(x/\ell)
\,\,,
\end{align*}
where $\ell$ is the healing length, Eq.~(\ref{healing_length}).
The potential seen by the CoM of the soliton scattered off a $\delta$-function barrier
\begin{align*}
&
v_{\text{barrier}}(x) = \tilde{g} \delta(x) 
\end{align*}
can be computed as 
\begin{align*}
&
V_{\text{soliton-on-barrier}} \stackrel{N\gg 1}{\approx} \int \! dx' \, n(x') v_{\text{barrier}}(x'+X)  =  \tilde{g} n(X)\,;
\end{align*}
see 
Refs.~\cite{weiss2009_010403,weiss2012_455306}.
This gives
\begin{align}
\max_{X} V_{\text{soliton-on-barrier}}(X) \stackrel{N\gg 1}{\approx}  \frac{1}{4} \frac{m |g| \tilde{g} N^2}{\hbar^2}
\label{barrier_top}
\,.
\end{align}

The ionization threshold \eqref{exact_ionization_threshold} and the single-atom ionization window \eqref{exact_single_ionization_window} respectively become
\begin{align}
E_{\text{kinetic, CoM}}  > \mu_{N}
\label{MF_ionization_threshold}
\,\,,
\end{align}
and 
\begin{align}
2\mu_{N} \ge E_{\text{kinetic, CoM}}  > \mu_{N}
\label{MF_single_ionization_window}
\,\,,
\end{align}
where 
\begin{align}
\mu_{N} = \frac{N^2}{8} \frac{mg^2}{\hbar^2}
\label{MF_mu}
\end{align}
is the chemical potential of an $N$-atom soliton, in the mean-field limit. For the 6-atom solitons used in this work, exact formulas must be used starting 
from about $3$-atom excitations.

\section{A complete interferometric cycle with a uniform field as a phase object\label{ssec_interferometer}}
\begin{figure}[h]
\centering
\includegraphics[width=.47\textwidth]{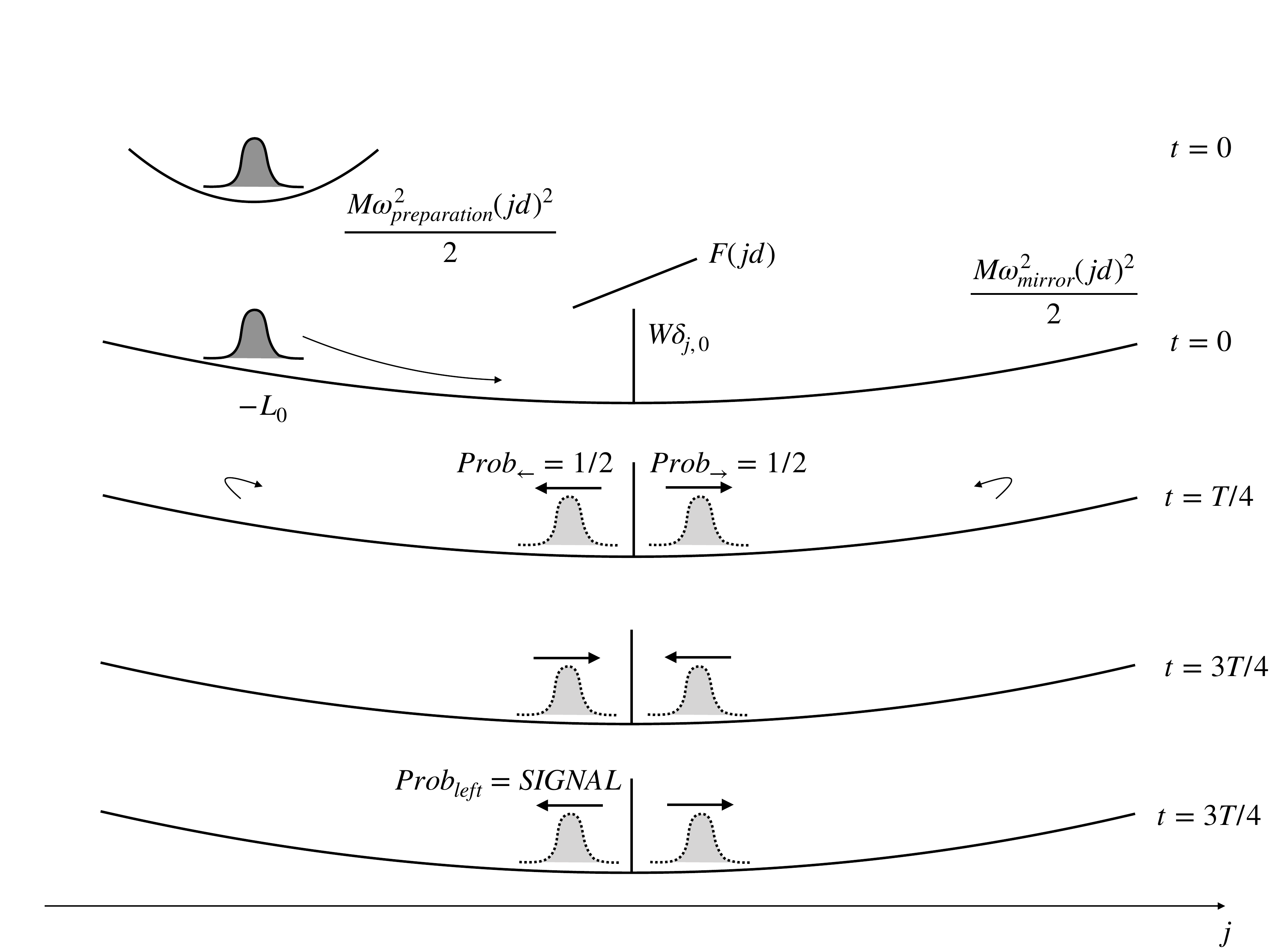}
\caption{
A complete interferometric cycle.
             }
\label{f:scheme}        
\end{figure}

Figure~\ref{f:scheme} shows a complete interferometric cycle. The atoms are prepared in the solitonic ground state of a harmonic oscillator of frequency $\omega_{\text{preparation}}$. At $t=0$, the soliton is released from its initial CoM position at $-L_{0}$, and at $t=T/4$, it hits the 
barrier ($T \equiv 2\pi/\omega_{\text{mirror}}$). The barrier is tuned to a 50\%--50\%  splitting. At $t=3T/4$, both wavepackets return to the barrier, where they interfere. We will be interested in the probability of 
finding the soliton to (say) the left of the barrier. This probability will be sensitive to the presence of a phase object, which will be represented by a  uniform force field of a (one-body) intensity $F$.
%
 %
%
%
\subsection{The prediction for the fringes, assuming an elastic scattering of the CoM off the barrier. }

 In the absence of inelastic effects, for a spatially even beamsplitter, the signal will behave as 
\begin{align}
&
\text{Prob}_{\text{left}}(F) = \sin^{2} \left(\frac{2 N F L_{0}}{\hbar\omega_{\text{mirror}}}\right)
\label{fringes}
\,;
\end{align}
we will derive this in the next subsection. Notice the ``quantum advantage factor'' $N$ appearing in the argument of the sine function. 

Due to velocity filtering, there appears a difference between the kinetic energies of the reflected (left) and transmitted (right) wavepackets. However, this does \emph{not} introduce any phase shift on recombination.

\subsection{A derivation of Eq.~(\ref{fringes}) for the signal absent inelastic effects }
Let us first define the phase $\phi_{\text{right}}$ as the \emph{total} phase accumulated by the right wavepacket of the CoM wavefunction between the 
beam-splttinng and recombination. It will not include the phase acquired in course of the beamsplitting process itself. The phase $\phi_{\text{right}}$ can be 
decomposed as 
\begin{align}
\phi_{\text{right}}^{} = \phi_{\text{right}}^{(0)} + \delta \phi_{\text{right}}^{}
\,.
\label{right_phase_decomposition}
\end{align}
Here $\phi_{\text{right}}^{(0)}$ is the phase that would be accumulated if the ``phase object'' were not present, and $\delta \phi_{\text{right}}$ is 
the contribution from the ``phase object.'' Anologously, we introduce 
\begin{align}
\phi_{\text{left}}^{} = \phi_{\text{left}}^{(0)} + \delta \phi_{\text{left}}^{}\,.
\label{left_phase_decomposition}
\end{align}
\subsubsection{The signal for a given phase difference $\phi_{\text{right}}-\phi_{\text{left}}$}
For our interferometer, the signal will be defined as the probability of finding the soliton to the left from the barrier after the recombination. 

We are assuming elastic scattering and no external potential besides the barrier. Then, asymptotically, the scattering solution $\psi_{\text{CoM}}(\bar{X}) $ for the CoM wave function  assumes the form 
\begin{align*}
\psi_{\text{CoM}}(\bar{X}) = 
\left\{
\begin{array}{ccc}
e^{+i\bar{K}\bar{X}} + r e^{-ik\bar{X}} & \text{ for } & \bar{X} \to -\infty
\\
t e^{+i\bar{K}\bar{X}}                 & \text{ for } & \bar{X} \to +\infty
\end{array}
\right.
\,\,,
\end{align*}
where $r$ and $t$ are respectively the reflection and transmission coefficients. 

It is easy to show that, after the beam-splitting and recombination, the signal has the form
\begin{align*}
\text{Prob}_{\text{left}} = |r^2 e^{i\phi_{\text{left}}} + t^2 e^{i\phi_{\text{right}}}|^2
\,.
\end{align*}
%

From the conservation of matter, we have $|r|^2 + |t|^2 = 1 $, so that, a priori, the family of possible values for $r$ and $t$ is parametrized by three real numbers. But if the scatterer used for both beam-splitting and recombination is spatially even, then $r$ and $t$  are more constrained and their possible values form a family parametrized by two real numbers, $\eta_{\text{e}}$ and $\eta_{\text{o}}$ (see, e.g.\ \cite{olshanii1998_938}). Here `e' stands for even and `o' for odd waves. This parametrization is as follows:
\begin{align*}
&
r = f_{\text{e}} - f_{\text{o}}
\\
&
t = 1 + f_{\text{e}} + f_{\text{o}}
\,\,,
\end{align*}
where
\begin{align*}
f_{\text{e,o}} = - \frac{1}{1+i\eta_{\text{e,o}}} 
\end{align*}
are the scattering amplitudes for the even and odd waves.   The signal now reads 
\begin{multline*}
\text{Prob}_{\text{left}} = \frac{1}{\left(\eta_{\text{e}}^2+1\right)^2\left(\eta_{\text{o}}^2+1\right)^2}
\\
\times \Big\{
-2 (\eta_{\text{e}} \eta_{\text{o}}+1)^2 (\eta_{\text{e}}-\eta_{\text{o}})^2 \cos \left(\phi_{\text{right}}-\phi_{\text{left}}\right)
\\
+(\eta_{\text{e}}-\eta_{\text{o}})^4+(\eta_{\text{e}} \eta_{\text{o}}+1)^4
\Big\}
\,\,.
\end{multline*}
We next require that the scatterer be a 50\%--50\% beam-splitter:
\begin{align*}
|t|^2 = |r|^2 = \frac{1}{2}
\,\,.
\end{align*}
There are two disjoint one-parametric families of $r$ and $t$ that have this property, and both can be parametrized by $\eta_{\text{e}}$:
\[
%
%
\eta_{\text{o}} = - \frac{\eta_{\text{e}} +1}{\eta_{\text{e}}-1}
\quad\text{and}\quad
\eta_{\text{o}} = + \frac{\eta_{\text{e}} -1}{\eta_{\text{e}}+1}
\,.
%
%
\]
For both families, $\eta_{\text{e}}$ can be any real number except $1$ for the first family and $-1$ for the second.
While the magnitudes of $r$ and $t$ are now fixed to 1/2, their phases will depend on the choice of family and the choice of the value of $\eta_{\text{e}}$.  Nonetheless, unexpectedly and inexplicably (in the sense that we don't know of any a priori reason why the mathematics had to work out this way), in an interferometer featuring a 50\%--50\% beam-splitter and a 50\%--50\% recombiner, the produced signal obeys a universal 
formula that depends \emph{only} on the phases accumulated between beam-splitting and recombination (and not on the phases of $r$ and $t$, i.e.\ neither on the choice of the family nor on the choice of $\eta_{\text{e}}$): 
\begin{align}
\text{Prob}_{\text{left}} = \sin^{2}
\left(
\frac{\phi_{\text{right}}-\phi_{\text{left}}}{2}
\right)\,.
\label{fringes_general}
\end{align}
To reiterate, this formula applies to any spatially \emph{even} scatterer, which must be \emph{the same} for both beam-splitting and recombination.  

\subsubsection{The vanishing of the unperturbed phase shift, $\phi_{\text{\rm right}}^{(0)} - \phi_{\text{\rm left}}^{(0)}$}
First, let us discuss the effect of the velocity filtering. There will be a difference in kinetic energies between the slow part of the incident wavepacket that gets 
reflected (left interferometer arm) by the barrier and the fast component that is transmitted  (right interferometer arm). A priori, this difference is expected to introduce a phase shift between the arms on recombination. This phase shift will depend on both the width and the shape of the velocity distribution of the incident wavepacket. Moreover, since the ``slow'' and the ``fast'' trajectories arriving at the recombiner at the same time would have left leave the beamspitter at two different 
instances of time, our interferometer would require a degree of spatio-temporal  coherence. Notice, however, that for an interferometer formed by a 
\emph{harmonic potential}, the latter effect disappears. This is not a coincidence, but an indication that in a harmonic interferometer, velocity filtering 
does \emph{not} introduce any additional left-right arm phase shift at all. This is indeed the case and can be proven in three ways: 
(a) quantum-mechanically, (b) semi-classically, using an explicit calculation, and (c) semi-classically, using a variational principle, for small energy differences between the arms only. 
Moreover, each of the two phases, 
$\phi_{\text{\rm left}}^{(0)}$ and $\phi_{\text{\rm right}}^{(0)}$, vanish separately. 
\begin{itemize}
\item[(a)] 
In a harmonic potential, any initial state \mbox{$\psi(x,\,t\!=\!0)$} gets transformed, after a half-period, to a state \mbox{$\psi\left(x,\,t\!=\!\pi/\omega_{\text{mirror}}\right) = -i \psi(-x,\,t\!=\!0)$}, 
i.e.\ to a mirror image of the initial state. No energy-dependent effects are present.
\item[(b)] 
Semi-classically, the phase acquired by the right wavepacket between beam-splitting and recombination will be given by the classical action:
\begin{align*}
\phi_{\text{\rm right}}^{(0)}
&= S_{\text{\rm right}}^{(0)}/\hbar
\\
&=\frac{1}{\hbar} \int_{t^{\text{BS}}}^{t^{\text{REC}}=t^{\text{BS}}+T/2} \!dt \, \mathcal{L}(\bar{X}_{\text{\rm right}}^{(0)}(t),\,\bar{\mathcal{V}}_{\text{\rm right}}^{(0)}(t))
\\
 &= \frac{1}{\hbar}\int_{t^{\text{BS}}}^{t^{\text{REC}}=t^{\text{BS}}+T/2} \!dt \, 
\left\{
\frac{1}{2}M \bar{\mathcal{V}}_{0}^2 \cos^2(\omega_{\text{mirror}} t)
\right.
\\
&
\hspace{5em}\left.
- 
\frac{1}{2}m \omega_{\text{mirror}}^2(\bar{\mathcal{V}}_{0}/\omega_{\text{mirror}})^2\sin^2(\omega_{\text{mirror}} t)\right\}
\\
&=\frac{1}{\hbar} \frac{E_{\text{\rm right}}}{\omega_{\text{mirror}}}  \int_{t^{\text{BS}}/\omega_{\text{mirror}}}^{t^{\text{BS}}/\omega_{\text{mirror}}+\pi} \!d\xi \, \left\{\cos^2(\xi) -\sin^2(\xi)
\right\}
\\
&=0
\,\,,
\end{align*}
where $\mathcal{L}(\bar{X},\,\bar{\mathcal{V}}) \equiv  M \bar{\mathcal{V}}^2/2 - M \omega_{\text{mirror}}^2\bar{X}^2/2$ is the classical Lagrangian for 
the CoM motion,  
$$
\bar{X}_{\text{\rm right}}^{(0)}(t) = (\bar{\mathcal{V}}_{0}/\omega_{\text{mirror}}) \sin(\omega_{\text{mirror}} t)
$$ 
and 
$$
\bar{\mathcal{V}}_{\text{\rm right}}^{(0)}(t) = \bar{\mathcal{V}}_{0} \cos(\omega_{\text{mirror}} t)
$$ are the unperturbed CoM trajectory and the corresponding velocity 
dependence in the right arm, and $M \equiv m N$ is the soliton mass. Further, 
\[\bar{\mathcal{V}}_{0} = L_{0} \omega_{\text{mirror}}\] 
is the velocity of the wavepacket on the beamsplitter, $E_{\text{\rm right}} = M \bar{\mathcal{V}}_{0}^2/2$ is its 
energy, and $t^{\text{BS}}$ and $t^{\text{REC}}$ are the beamsplitting and recombination time instances. 
As one can see, the action  \emph{vanishes identically} for any energy of the wavepacket. This derivation can be repeated verbatim for the left interferometer arm. This proof shows that $\phi_{\text{\rm left}}^{(0)} = 0$ and $\phi_{\text{\rm right}}^{(0)} = 0$ separately. 
\item[(c)]
Finally, the absence of the energy dependence of the phase accumulated between the beamsplitting and recombination can be proven variationally, 
within the semicalssical approximation, for small energy variations. Notice, again, that the slow (left) and the fast (right) wavepackets share the initial and the final points  of 
their trajectories, in \emph{both} space and time. If the energy difference between the trajectories is small, then one of them can be considered a small  variation of the other, with fixed space-time end-points. Since the latter trajectory obeys laws of classical mechanics, it must obey principle of least action. Thus, the difference 
between the two actions (hence between the two quantum phases, in the semi-classical approximation) must vanish to linear order in the amplitude of trajectory variation.   
\end{itemize}
To sum up, we have just showed that 
\begin{align}
\phi_{\text{right}}^{(0)} - \phi_{\text{left}}^{(0)} = 0 
\,\,.
\label{principal_phase_shift}
\end{align}
This property is specific for an interferometric cycle driven by a harmonic potential. Since the right-left energy disparity is conjectured to be 
unavoidable in interferometry with massive objects \cite{weiss2009_010403,weiss2012_455306}, a harmonic
control of the interferometer arms may provide a remedy for a possible dependence of the
fringe position on the energy and shape of the wavepacket.  
\subsubsection{An explicit calculation of the fringe shift due to a uniform field}
Finally, we turn to the phase shift accumulated due to the ``phase object.'' Let us compute the phase shift induced by the uniform field $Fx$ on, for example, the right arm of the interferometer. 
Note that the potential energy correction $Fx$ refers to a single atom. For the CoM, the energy correction \emph{and the resulting quantum phase correction accumulated} must both be multiplied by the number of atoms $N$. The resulting correction to the CoM Lagrangian becomes 
\[
\delta \mathcal{L}(\bar{X},\,\bar{\mathcal{V}}) = -NF\bar{X}
\,.
\]
This amplification, combined with a suppression (which is much harder to achieve) of decoherence of the CoM motion to other degrees of freedom, paves the way to \emph{quantum advantage in particle interferometry}.

Using the principle of the least action, one can easily show (see, e.g.\ Ref.~\cite{olshanii1994_995}) that a correction to 
the arm trajectory introduced by a correction to the Lagrangian does not contribute to a correction to the acton, in the first 
order in $\delta \mathcal{L}$. Thus, 
\begin{align*}
\delta\phi_{\text{\rm right}}
&= \delta S_{\text{\rm right}}/\hbar
\\
&= \frac{1}{\hbar}\int_{t^{\text{BS}}}^{t^{\text{REC}}=t^{\text{BS}}+T/2} \!dt \, \delta\mathcal{L}(\bar{X}_{\text{\rm right}}(t),\,\bar{\mathcal{V}}_{\text{\rm right}}(t))
\\
&=\frac{1}{\hbar} \int_{t^{\text{BS}}}^{t^{\text{REC}}=t^{\text{BS}}+T/2} \!dt \, 
(-NF L_{0})\sin(\omega_{\text{mirror}} t)
\\
&= 
-\frac{NF L_{0}}{\hbar\omega_{\text{mirror}}} 
\int_{t^{\text{BS}}/\omega_{\text{mirror}}}^{t^{\text{BS}}/\omega_{\text{mirror}}+\pi} \!d\xi \, \sin(\xi)
\\
&=
-\frac{2 NF L_{0}}{\hbar\omega_{\text{mirror}}} 
\,\,.
\end{align*}
An analogous computation for the left arm gives 
\[\delta\phi_{\text{\rm left}} = + \frac{2 NF L_{0}}{\hbar\omega_{\text{mirror}}}\,.
\]

Putting the two contributions together, we get 
\begin{align}
\delta\phi_{\text{right}} - \delta\phi_{\text{left}} =-\frac{4 NF L_{0}}{\hbar\omega_{\text{mirror}}}  
\,\,.
\label{phase_shift_correction}
\end{align}
%

%
%
Combining our results in \cref{right_phase_decomposition,left_phase_decomposition,fringes_general,principal_phase_shift,phase_shift_correction}, 
we finally obtain Eq.~\eqref{fringes}.

\section{Numerical simulations \label{sec_runs}}
%
Our principal interest was to investigate how the degree to which soliton `ionization' is energetically allowed (i.e.\ how large is the CoM kinetic energy of the soliton, Eq.~\eqref{CoM_kinetic_energy}, as compared to the various thresholds in Sec.~\ref{sec:inelastic_effects}) affects the degradation of interference fringes from their idealized behavior in Eq.~\eqref{fringes}. 

The ratio of the CoM kinetic energy to the ionization thresholds can be tuned by adjusting one or more of $\kappa_{\text{mirror}}$, $U$, and $W$. Once the parameters that determine this ratio are set, we compute $\text{Prob}_{\text{left}}$ (in fact, simply the number of atoms $N_{\text{L}}$ detected to the left of the barrier, since $\text{Prob}_{\text{left}}=N_{\text{L}}/N$) for various values of $F$. This produces a curve like that in Fig.~\ref{f:DATI_6_fringes}, featuring interference fringes which are a function of $F$. We will present two such curves, corresponding to two different ratios of the CoM kinetic energy to the ionization threshold.

We work in a system of units in which
%
%
%
\(
 d=J=\hbar=1\,.
\)
%
In all the runs that we will present, the number of atoms is $N=6$, the number of lattice sites is $L=29$, and the initial offset is $L_{0}=7$. In principle, the condition in Eq.~\eqref{trap_frequency_to_incident_E} links the values of $\kappa_{\text{preparation}}$, $U$, $W$, so that only two of them can be chosen independently. However, after some experimentation, we found that best fringes are obtained if this condition is slightly violated. We ended up varying just the value of $U$ while keeping $\kappa_{\text{preparation}} = 0.00002813$ and  $W=0.24$. Furthermore,  we also kept $\kappa_{\text{mirror}} = 0.001800$. 

Under these conditions, we can vary the ratio of the CoM kinetic energy to the ionization thresholds by varying $U$. For each choice of $U$, need to produce a curve of $\text{Prob}_{\text{left}}$ versus $F$. We do this by picking a range of values of $F$; for each such value, we numerically simulate, using exact diagonalization, the complete interferometric cycle from Fig.~\ref{f:scheme}, and compute the number of atoms $N_{\text{L}}$ detected to the left of the barrier at the end of the cycle.

We should note that we ended up slightly modifying the interferometric cycle in Fig.~\ref{f:scheme}, in the folllowing way: instead of having the field $F$ be present for the entire cycle, we only `turn it on' slightly before the soliton first hits the barrier (at $t=17.9$ time units), and then `turn it off' slightly after the recombination (at $t=46.5$ time units). The reason we did this is that the influence of the $F$ field on the trajectory of the solitons is not completely negligible, and, at the recombining stage, the two wavepackets arrive at the beamsplitter at somewhat diferent times. This effect is not accounted for at all in our idealized theory that leads to  Eq.~\eqref{fringes}. Indeed, for a sufficiently strong $F$, these times of arrival will be so different that the incoming wavepackets will never overlap atop the beamsplitter: they will completely miss each other and there will be no recombination. In order to minimize this effect, we keep $F$ `turned off' for as long as possible at the beginning and at the end of the cycle.

Having obtaned a numerical curve of $N_{\text{L}}$ versus $F$, we compare it to the idealized result in Eq.~\eqref{fringes} by fitting the numerical curve to the following function:
\begin{multline*}
 y(F)=a_{1}+a_{2} F+a_{3} F^{2}
 \\
 +\left(b_{1}+b_{2} F+b_{3} F^{2}\right)\left[\cos^{2}\left(\omega_{\text{f\/it}} F + \phi_{0}\right) - \frac{1}{2}\right]\,.
\end{multline*}
There are 8 fitting parameters ($a_{1},\,a_{2},\,a_{3},\,b_{1},\,b_{2},\,b_{3},\,\omega_{\text{f\/it}},\,$ and $\phi_{0}$), but in reality we are only interested in $\omega_{\text{f\/it}}$. According to Eq.~\eqref{fringes}, $\omega_{\text{f\/it}}$ should be correspond to $\frac{2 N L_{0}}{\hbar\omega_{\text{mirror}}}$, which works out to $990$ for our chosen values of parameters. All the other fitting parameters are merely `empirical', introduced to account for deviations from the idealized behavior in Eq.~\eqref{fringes}.

\subsection{Results}
\begin{figure*}
    \mbox{}\hfill
    \begin{minipage}{0.45\textwidth}
        \centering
        \includegraphics[width=0.9\textwidth]{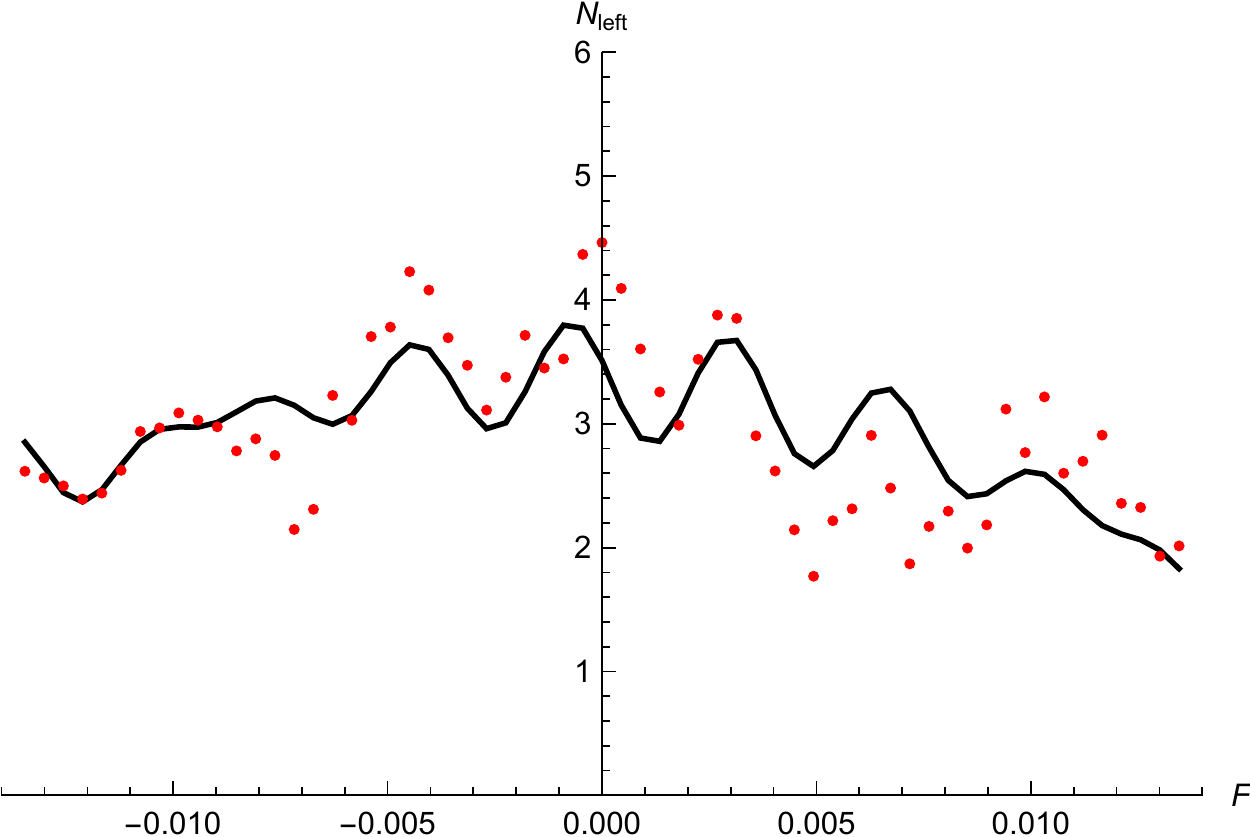}
        \caption{Interference fringes for $U=-0.3$. The CoM kinetic energy is sufficiently high that total ionization is energetically allowed.
The fitted value of the frequency of oscillations is $\omega_{\text{f\/it}}=846$.
}
\label{f:DATI_6_fringes} 
    \end{minipage}\hfill
    \begin{minipage}{0.45\textwidth}
        \centering
        \includegraphics[width=0.9\textwidth]{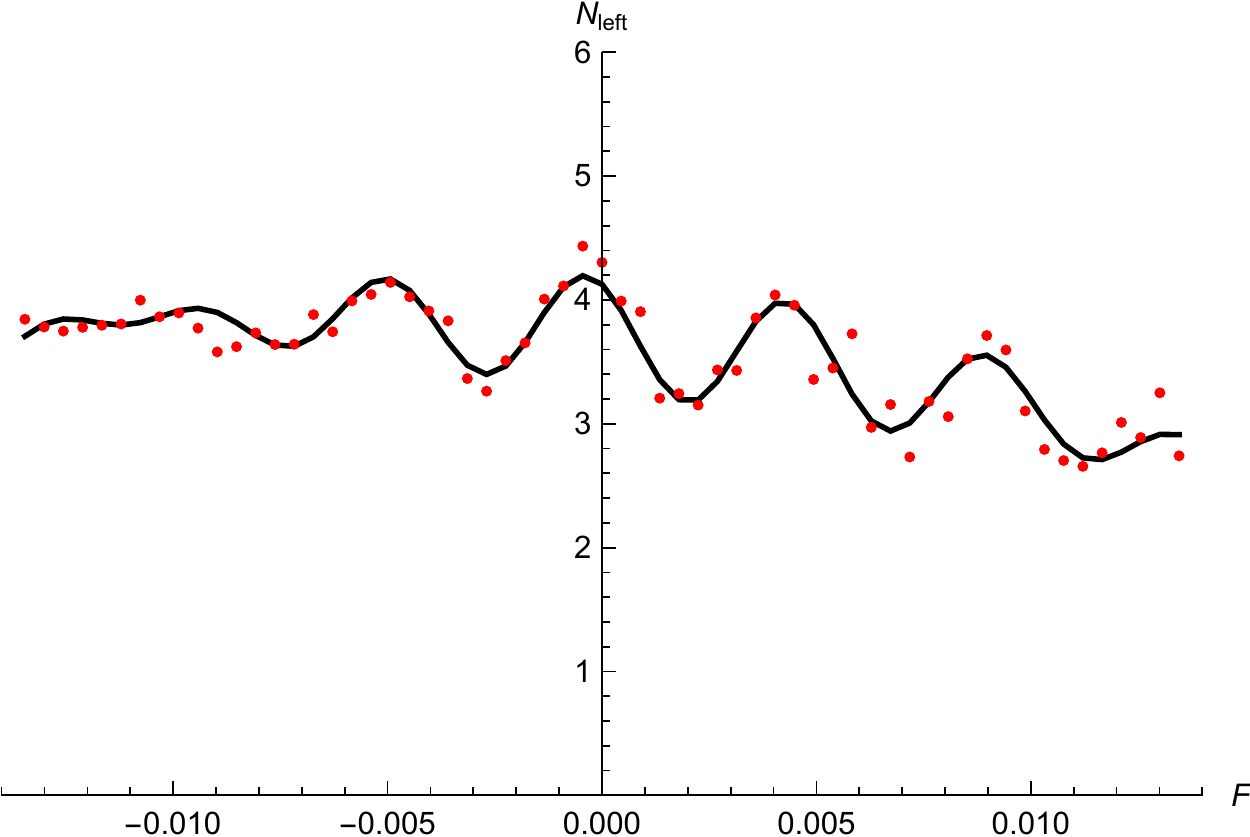}
        \caption{Interference fringes for $U=-0.4$. The CoM kinetic energy is just above the threshold necessary in order for shedding of two atoms (two-particle ionization) to be energetically allowed.
The fitted value of the frequency of oscillations is $\omega_{\text{f\/it}}=667$. 
}
\label{f:DATI_7_fringes} 
    \end{minipage}
    \hfill\mbox{}
\end{figure*}

Our results are presented in Figs.~\ref{f:DATI_6_fringes}
and \ref{f:DATI_7_fringes}, corresponding respectively to the choices $U=-0.3$ and $U=-0.4$.

A priori, our expectation is that $U = -0.3$ allows for a complete disintegration of the soliton onto six individual atoms, so a significant suppression of fringes is expected. On the other hand, $U = -0.4$ brings us slightly above the  double ionization: we have enough energy to extract two atoms from the total of six, but no more. 
Thus, some suppression of fringes is expected, but less than for $U = -0.3$.

It is not clear if the numerical results bare out these expectations. From the figures, it seems that the $U = -0.3$ plot is noisier than the $U = -0.4$ one, but it is not obvious that the fringes are more suppressed for $U = -0.3$. Moreover, the fitted value of the frequency is closer to the idealized one (990) for $U = -0.3$ than for $U = -0.4$.

Our main result is that the interferometric signal still exists even when a complete disintegration of the soliton onto six individual atoms is energetically allowed.

%
\section{Conclusions}

We simulated the matter-wave interferometer of the kind originally proposed by Weiss and Castin \cite{weiss2009_010403,weiss2012_455306}. Unlike in the original proposal, we considered situations where either partial or complete disintegration of the soliton is energetically allowed. Our expectation was that the suppression of the interferometric fringes will depend in some simple way on the number of atoms that the soliton is energetically allowed to lose. The numerical simulations did not bare this out: the fringe distorsion sequence is more complicated, and further study is required to make sense of it. Our main result is that, surprisingly, the interferometric signal survives even when the soliton has enough energy to completely disintegrate.

\section{Acknowledgments}

We benefited from useful discssions with Luigi Amico.\\

This work was supported by the NSF grants PHY-1912542 and PHY-1607221.

\bibliography{Nonlinear_PDEs_and_SUSY_v038,Bethe_ansatz_v033,anomaly_v017,thermalization_literature_v040,soliton_interferometry_01}

\begin{thebibliography}{19}%
\makeatletter
\providecommand \@ifxundefined [1]{%
 \@ifx{#1\undefined}
}%
\providecommand \@ifnum [1]{%
 \ifnum #1\expandafter \@firstoftwo
 \else \expandafter \@secondoftwo
 \fi
}%
\providecommand \@ifx [1]{%
 \ifx #1\expandafter \@firstoftwo
 \else \expandafter \@secondoftwo
 \fi
}%
\providecommand \natexlab [1]{#1}%
\providecommand \enquote  [1]{``#1''}%
\providecommand \bibnamefont  [1]{#1}%
\providecommand \bibfnamefont [1]{#1}%
\providecommand \citenamefont [1]{#1}%
\providecommand \href@noop [0]{\@secondoftwo}%
\providecommand \href [0]{\begingroup \@sanitize@url \@href}%
\providecommand \@href[1]{\@@startlink{#1}\@@href}%
\providecommand \@@href[1]{\endgroup#1\@@endlink}%
\providecommand \@sanitize@url [0]{\catcode `\\12\catcode `\$12\catcode
  `\&12\catcode `\#12\catcode `\^12\catcode `\_12\catcode `\%12\relax}%
\providecommand \@@startlink[1]{}%
\providecommand \@@endlink[0]{}%
\providecommand \url  [0]{\begingroup\@sanitize@url \@url }%
\providecommand \@url [1]{\endgroup\@href {#1}{\urlprefix }}%
\providecommand \urlprefix  [0]{URL }%
\providecommand \Eprint [0]{\href }%
\providecommand \doibase [0]{https://doi.org/}%
\providecommand \selectlanguage [0]{\@gobble}%
\providecommand \bibinfo  [0]{\@secondoftwo}%
\providecommand \bibfield  [0]{\@secondoftwo}%
\providecommand \translation [1]{[#1]}%
\providecommand \BibitemOpen [0]{}%
\providecommand \bibitemStop [0]{}%
\providecommand \bibitemNoStop [0]{.\EOS\space}%
\providecommand \EOS [0]{\spacefactor3000\relax}%
\providecommand \BibitemShut  [1]{\csname bibitem#1\endcsname}%
\let\auto@bib@innerbib\@empty
\bibitem [{\citenamefont {Cuevas}\ \emph {et~al.}(2013)\citenamefont {Cuevas},
  \citenamefont {Kevrekidis}, \citenamefont {Malomed}, \citenamefont {Dyke},\
  and\ \citenamefont {Hulet}}]{cuevas2013_063006}%
  \BibitemOpen
  \bibfield  {author} {\bibinfo {author} {\bibfnamefont {J.}~\bibnamefont
  {Cuevas}}, \bibinfo {author} {\bibfnamefont {P.~G.}\ \bibnamefont
  {Kevrekidis}}, \bibinfo {author} {\bibfnamefont {B.~A.}\ \bibnamefont
  {Malomed}}, \bibinfo {author} {\bibfnamefont {P.}~\bibnamefont {Dyke}},\ and\
  \bibinfo {author} {\bibfnamefont {R.~G.}\ \bibnamefont {Hulet}},\ }\bibfield
  {title} {\bibinfo {title} {Interactions of solitons with a {G}aussian
  barrier: splitting and recombination in quasi-one-dimensional and
  three-dimensional settings},\ }\href@noop {} {\bibfield  {journal} {\bibinfo
  {journal} {New J. Phys.}\ }\textbf {\bibinfo {volume} {15}},\ \bibinfo
  {pages} {063006} (\bibinfo {year} {2013})}\BibitemShut {NoStop}%
\bibitem [{\citenamefont {Veretenov}\ \emph {et~al.}(2007)\citenamefont
  {Veretenov}, \citenamefont {Rozhdestvensky}, \citenamefont {Rosanov},
  \citenamefont {Smirnov},\ and\ \citenamefont {Fedorov}}]{Veretenov2007_455}%
  \BibitemOpen
  \bibfield  {author} {\bibinfo {author} {\bibfnamefont {N.}~\bibnamefont
  {Veretenov}}, \bibinfo {author} {\bibfnamefont {Y.}~\bibnamefont
  {Rozhdestvensky}}, \bibinfo {author} {\bibfnamefont {N.}~\bibnamefont
  {Rosanov}}, \bibinfo {author} {\bibfnamefont {V.}~\bibnamefont {Smirnov}},\
  and\ \bibinfo {author} {\bibfnamefont {S.}~\bibnamefont {Fedorov}},\
  }\bibfield  {title} {\bibinfo {title} {Interferometric precision measurements
  with {B}ose-{E}instein condensate solitons formed by an optical lattice},\
  }\href {https://doi.org/10.1140/epjd/e2007-00129-2} {\bibfield  {journal}
  {\bibinfo  {journal} {Eur. Phys. J. D}\ }\textbf {\bibinfo {volume} {42}},\
  \bibinfo {pages} {455} (\bibinfo {year} {2007})}\BibitemShut {NoStop}%
\bibitem [{\citenamefont {Streltsov}\ \emph
  {et~al.}(2009{\natexlab{a}})\citenamefont {Streltsov}, \citenamefont {Alon},\
  and\ \citenamefont {Cederbaum}}]{streltsov2009_043616}%
  \BibitemOpen
  \bibfield  {author} {\bibinfo {author} {\bibfnamefont {A.~I.}\ \bibnamefont
  {Streltsov}}, \bibinfo {author} {\bibfnamefont {O.~E.}\ \bibnamefont
  {Alon}},\ and\ \bibinfo {author} {\bibfnamefont {L.~S.}\ \bibnamefont
  {Cederbaum}},\ }\bibfield  {title} {\bibinfo {title} {Scattering of an
  attractive {B}ose-{E}instein condensate from a barrier: {F}ormation of
  quantum superposition states},\ }\href@noop {} {\bibfield  {journal}
  {\bibinfo  {journal} {Phys. Rev. A}\ }\textbf {\bibinfo {volume} {80}},\
  \bibinfo {pages} {043616} (\bibinfo {year} {2009}{\natexlab{a}})}\BibitemShut
  {NoStop}%
\bibitem [{\citenamefont {Streltsov}\ \emph
  {et~al.}(2009{\natexlab{b}})\citenamefont {Streltsov}, \citenamefont {Alon},\
  and\ \citenamefont {Cederbaum}}]{Streltsov2009_091004}%
  \BibitemOpen
  \bibfield  {author} {\bibinfo {author} {\bibfnamefont {A.~I.}\ \bibnamefont
  {Streltsov}}, \bibinfo {author} {\bibfnamefont {O.~E.}\ \bibnamefont
  {Alon}},\ and\ \bibinfo {author} {\bibfnamefont {L.~S.}\ \bibnamefont
  {Cederbaum}},\ }\bibfield  {title} {\bibinfo {title} {Efficient generation
  and properties of mesoscopic quantum superposition states in an attractive
  {B}ose-{E}instein condensate threaded by a potential barrier},\ }\href
  {https://doi.org/10.1088/0953-4075/42/9/091004} {\bibfield  {journal}
  {\bibinfo  {journal} {J. Phys. Bs}\ }\textbf {\bibinfo {volume} {42}},\
  \bibinfo {pages} {091004} (\bibinfo {year} {2009}{\natexlab{b}})}\BibitemShut
  {NoStop}%
\bibitem [{\citenamefont {Weiss}\ and\ \citenamefont
  {Castin}(2009)}]{weiss2009_010403}%
  \BibitemOpen
  \bibfield  {author} {\bibinfo {author} {\bibfnamefont {C.}~\bibnamefont
  {Weiss}}\ and\ \bibinfo {author} {\bibfnamefont {Y.}~\bibnamefont {Castin}},\
  }\bibfield  {title} {\bibinfo {title} {Creation and detection of a mesoscopic
  gas in a nonlocal quantum superposition},\ }\href@noop {} {\bibfield
  {journal} {\bibinfo  {journal} {Phys. Rev. Lett.}\ }\textbf {\bibinfo
  {volume} {102}},\ \bibinfo {pages} {010403} (\bibinfo {year}
  {2009})}\BibitemShut {NoStop}%
\bibitem [{\citenamefont {Naldesi}\ \emph {et~al.}(2020)\citenamefont
  {Naldesi}, \citenamefont {Gomez}, \citenamefont {Dunjko}, \citenamefont
  {Perrin}, \citenamefont {Olshanii}, \citenamefont {Amico},\ and\
  \citenamefont {Minguzzi}}]{naldesi2020enhancing}%
  \BibitemOpen
  \bibfield  {author} {\bibinfo {author} {\bibfnamefont {P.}~\bibnamefont
  {Naldesi}}, \bibinfo {author} {\bibfnamefont {J.~P.}\ \bibnamefont {Gomez}},
  \bibinfo {author} {\bibfnamefont {V.}~\bibnamefont {Dunjko}}, \bibinfo
  {author} {\bibfnamefont {H.}~\bibnamefont {Perrin}}, \bibinfo {author}
  {\bibfnamefont {M.}~\bibnamefont {Olshanii}}, \bibinfo {author}
  {\bibfnamefont {L.}~\bibnamefont {Amico}},\ and\ \bibinfo {author}
  {\bibfnamefont {A.}~\bibnamefont {Minguzzi}},\ }\href@noop {} {\bibinfo
  {title} {Enhancing sensitivity to rotations with quantum solitonic currents}}
  (\bibinfo {year} {2020}),\ \Eprint {https://arxiv.org/abs/1901.09398}
  {arXiv:1901.09398 [cond-mat.quant-gas]} \BibitemShut {NoStop}%
\bibitem [{\citenamefont {Polo}\ \emph {et~al.}(2021)\citenamefont {Polo},
  \citenamefont {Naldesi}, \citenamefont {Minguzzi},\ and\ \citenamefont
  {Amico}}]{Polo2021_015015}%
  \BibitemOpen
  \bibfield  {author} {\bibinfo {author} {\bibfnamefont {J.}~\bibnamefont
  {Polo}}, \bibinfo {author} {\bibfnamefont {P.}~\bibnamefont {Naldesi}},
  \bibinfo {author} {\bibfnamefont {A.}~\bibnamefont {Minguzzi}},\ and\
  \bibinfo {author} {\bibfnamefont {L.}~\bibnamefont {Amico}},\ }\bibfield
  {title} {\bibinfo {title} {The quantum solitons atomtronic interference
  device},\ }\href {https://doi.org/10.1088/2058-9565/ac39f6} {\bibfield
  {journal} {\bibinfo  {journal} {Quantum Sci. Technol.}\ }\textbf {\bibinfo
  {volume} {7}},\ \bibinfo {pages} {015015} (\bibinfo {year}
  {2021})}\BibitemShut {NoStop}%
\bibitem [{\citenamefont {Naldesi}\ \emph {et~al.}(2019)\citenamefont
  {Naldesi}, \citenamefont {Gomez}, \citenamefont {Malomed}, \citenamefont
  {Olshanii}, \citenamefont {Minguzzi},\ and\ \citenamefont
  {Amico}}]{naldesi2018_053001}%
  \BibitemOpen
  \bibfield  {author} {\bibinfo {author} {\bibfnamefont {P.}~\bibnamefont
  {Naldesi}}, \bibinfo {author} {\bibfnamefont {J.~P.}\ \bibnamefont {Gomez}},
  \bibinfo {author} {\bibfnamefont {B.}~\bibnamefont {Malomed}}, \bibinfo
  {author} {\bibfnamefont {M.}~\bibnamefont {Olshanii}}, \bibinfo {author}
  {\bibfnamefont {A.}~\bibnamefont {Minguzzi}},\ and\ \bibinfo {author}
  {\bibfnamefont {L.}~\bibnamefont {Amico}},\ }\bibfield  {title} {\bibinfo
  {title} {Rise and fall of a bright soliton in an optical lattice},\
  }\href@noop {} {\bibfield  {journal} {\bibinfo  {journal} {Phys. Rev. Lett.}\
  }\textbf {\bibinfo {volume} {122}},\ \bibinfo {pages} {053001} (\bibinfo
  {year} {2019})}\BibitemShut {NoStop}%
\bibitem [{\citenamefont {Grimshaw}\ \emph {et~al.}(2021)\citenamefont
  {Grimshaw}, \citenamefont {Billam},\ and\ \citenamefont
  {Gardiner}}]{grimshaw2021soliton}%
  \BibitemOpen
  \bibfield  {author} {\bibinfo {author} {\bibfnamefont {C.~L.}\ \bibnamefont
  {Grimshaw}}, \bibinfo {author} {\bibfnamefont {T.~P.}\ \bibnamefont
  {Billam}},\ and\ \bibinfo {author} {\bibfnamefont {S.~A.}\ \bibnamefont
  {Gardiner}},\ }\href@noop {} {\bibinfo {title} {Soliton interferometry with
  very narrow barriers obtained from spatially dependent dressed states}}
  (\bibinfo {year} {2021}),\ \Eprint {https://arxiv.org/abs/2104.11511}
  {arXiv:2104.11511 [cond-mat.quant-gas]} \BibitemShut {NoStop}%
\bibitem [{\citenamefont {Holdaway}\ \emph {et~al.}(2012)\citenamefont
  {Holdaway}, \citenamefont {Weiss},\ and\ \citenamefont
  {Gardiner}}]{Holdaway2012_053618}%
  \BibitemOpen
  \bibfield  {author} {\bibinfo {author} {\bibfnamefont {D.~I.~H.}\
  \bibnamefont {Holdaway}}, \bibinfo {author} {\bibfnamefont {C.}~\bibnamefont
  {Weiss}},\ and\ \bibinfo {author} {\bibfnamefont {S.~A.}\ \bibnamefont
  {Gardiner}},\ }\bibfield  {title} {\bibinfo {title} {Quantum theory of bright
  matter-wave solitons in harmonic confinement},\ }\href
  {https://doi.org/10.1103/PhysRevA.85.053618} {\bibfield  {journal} {\bibinfo
  {journal} {Phys. Rev. A}\ }\textbf {\bibinfo {volume} {85}},\ \bibinfo
  {pages} {053618} (\bibinfo {year} {2012})}\BibitemShut {NoStop}%
\bibitem [{\citenamefont {Weiss}\ and\ \citenamefont
  {Castin}(2012)}]{weiss2012_455306}%
  \BibitemOpen
  \bibfield  {author} {\bibinfo {author} {\bibfnamefont {C.}~\bibnamefont
  {Weiss}}\ and\ \bibinfo {author} {\bibfnamefont {Y.}~\bibnamefont {Castin}},\
  }\bibfield  {title} {\bibinfo {title} {Elastic scattering of a quantum
  matter-wave bright soliton on a barrier},\ }\href@noop {} {\bibfield
  {journal} {\bibinfo  {journal} {J. Phys. A}\ }\textbf {\bibinfo {volume}
  {45}},\ \bibinfo {pages} {455306} (\bibinfo {year} {2012})}\BibitemShut
  {NoStop}%
\bibitem [{\citenamefont {Boiss\'{e}}\ \emph {et~al.}(2017)\citenamefont
  {Boiss\'{e}}, \citenamefont {Berthet}, \citenamefont {Fouch\'{e}},
  \citenamefont {Salomon}, \citenamefont {Aspect}, \citenamefont {Lepoutre},\
  and\ \citenamefont {Bourdel}}]{Boisse2017_10007}%
  \BibitemOpen
  \bibfield  {author} {\bibinfo {author} {\bibfnamefont {A.}~\bibnamefont
  {Boiss\'{e}}}, \bibinfo {author} {\bibfnamefont {G.}~\bibnamefont {Berthet}},
  \bibinfo {author} {\bibfnamefont {L.}~\bibnamefont {Fouch\'{e}}}, \bibinfo
  {author} {\bibfnamefont {G.}~\bibnamefont {Salomon}}, \bibinfo {author}
  {\bibfnamefont {A.}~\bibnamefont {Aspect}}, \bibinfo {author} {\bibfnamefont
  {S.}~\bibnamefont {Lepoutre}},\ and\ \bibinfo {author} {\bibfnamefont
  {T.}~\bibnamefont {Bourdel}},\ }\bibfield  {title} {\bibinfo {title}
  {Nonlinear scattering of atomic bright solitons in disorder},\ }\href@noop {}
  {\bibfield  {journal} {\bibinfo  {journal} {Eur. Phys. Lett.}\ }\textbf
  {\bibinfo {volume} {117}},\ \bibinfo {pages} {10007} (\bibinfo {year}
  {2017})}\BibitemShut {NoStop}%
\bibitem [{\citenamefont {Valiente}\ and\ \citenamefont
  {Petrosyan}(2008)}]{valiente2008_161002}%
  \BibitemOpen
  \bibfield  {author} {\bibinfo {author} {\bibfnamefont {M.}~\bibnamefont
  {Valiente}}\ and\ \bibinfo {author} {\bibfnamefont {D.}~\bibnamefont
  {Petrosyan}},\ }\bibfield  {title} {\bibinfo {title} {Two-particle states in
  the {H}ubbard model},\ }\href@noop {} {\bibfield  {journal} {\bibinfo
  {journal} {J. Phys. B}\ }\textbf {\bibinfo {volume} {41}},\ \bibinfo {pages}
  {161002} (\bibinfo {year} {2008})}\BibitemShut {NoStop}%
\bibitem [{\citenamefont {Castin}(2004)}]{castin2004_89}%
  \BibitemOpen
  \bibfield  {author} {\bibinfo {author} {\bibfnamefont {Y.}~\bibnamefont
  {Castin}},\ }\bibfield  {title} {\bibinfo {title} {Simple theoretical tools
  for low dimension {B}ose gases},\ }\href@noop {} {\bibfield  {journal}
  {\bibinfo  {journal} {J. Phys. IV France}\ }\textbf {\bibinfo {volume}
  {116}},\ \bibinfo {pages} {89} (\bibinfo {year} {2004})}\BibitemShut
  {NoStop}%
\bibitem [{\citenamefont {Castin}(2009)}]{castin2009_317}%
  \BibitemOpen
  \bibfield  {author} {\bibinfo {author} {\bibfnamefont {Y.}~\bibnamefont
  {Castin}},\ }\bibfield  {title} {\bibinfo {title} {Internal structure of a
  quantum soliton and classical excitations due to trap opening},\ }\href@noop
  {} {\bibfield  {journal} {\bibinfo  {journal} {Eur. Phys. J. B}\ }\textbf
  {\bibinfo {volume} {68}},\ \bibinfo {pages} {317} (\bibinfo {year}
  {2009})}\BibitemShut {NoStop}%
\bibitem [{\citenamefont {Calogero}\ and\ \citenamefont
  {Degasperis}(1975)}]{calogero1975_265}%
  \BibitemOpen
  \bibfield  {author} {\bibinfo {author} {\bibfnamefont {F.}~\bibnamefont
  {Calogero}}\ and\ \bibinfo {author} {\bibfnamefont {A.}~\bibnamefont
  {Degasperis}},\ }\bibfield  {title} {\bibinfo {title} {Comparison between the
  exact and {H}artree solutions of a one-dimensional many-body problem},\
  }\href@noop {} {\bibfield  {journal} {\bibinfo  {journal} {Phus. Rev. A}\
  }\textbf {\bibinfo {volume} {11}},\ \bibinfo {pages} {265} (\bibinfo {year}
  {1975})}\BibitemShut {NoStop}%
\bibitem [{Note1()}]{Note1}%
  \BibitemOpen
  \bibinfo {note} {It may seem that breathing excitations constitute another
  potential inelastic channel. However, curiously, the soliton does not possess
  true localized excitations: the mean-field breathers are, in reality,
  unbound. Interestingly, the absence of bound excitations is confirmed in the
  Bogoliubov approximation. Such a restoration may seem accidental, if
  Bogoliubov is to be considered as an approximation of mean-filed equations.
  However, as a linearization of the Heisenberg equations of motion for the
  quantum field, Bogoliubov can give predictions that are more accurate than
  the mean-field ones, albeit limited to (perhaps multiple) monomer
  excitations.}\BibitemShut {Stop}%
\bibitem [{\citenamefont {Olshanii}(1998)}]{olshanii1998_938}%
  \BibitemOpen
  \bibfield  {author} {\bibinfo {author} {\bibfnamefont {M.}~\bibnamefont
  {Olshanii}},\ }\bibfield  {title} {\bibinfo {title} {Atomic scattering in the
  presence of an external confinement and a gas of impenetrable bosons},\
  }\href@noop {} {\bibfield  {journal} {\bibinfo  {journal} {Phys. Rev. Lett.}\
  }\textbf {\bibinfo {volume} {81}},\ \bibinfo {pages} {938} (\bibinfo {year}
  {1998})}\BibitemShut {NoStop}%
\bibitem [{\citenamefont {Ol'shanii}(1994)}]{olshanii1994_995}%
  \BibitemOpen
  \bibfield  {author} {\bibinfo {author} {\bibfnamefont {M.}~\bibnamefont
  {Ol'shanii}},\ }\bibfield  {title} {\bibinfo {title} {Atomic
  interferometry},\ }\href@noop {} {\bibfield  {journal} {\bibinfo  {journal}
  {Laser Physics}\ }\textbf {\bibinfo {volume} {4}},\ \bibinfo {pages} {995}
  (\bibinfo {year} {1994})}\BibitemShut {NoStop}%
\end{thebibliography}%


\end{document}